\documentclass[epj]{svjour}

\usepackage{amsmath,amssymb,amsfonts}
\usepackage{graphicx}
\usepackage{hyperref}
\hypersetup{colorlinks=true,
            linkcolor=blue,
            citecolor=blue,
            urlcolor=blue}
\usepackage{xcolor}
\usepackage{booktabs}
\usepackage{cancel}
\usepackage{slashed}
\usepackage{microtype}
\usepackage{tikz}
\usepackage[compat=1.1.0]{tikz-feynman}
\usetikzlibrary{decorations.markings,arrows.meta}

\newcommand{\None}{N_1}
\newcommand{\Ntwo}{N_2}
\newcommand{\etapm}{\tilde{\eta}^{\pm}}
\newcommand{\etaz}{\tilde{\eta}^{0}}
\newcommand{\Aone}{A_1^0}

\newcommand{\ctauN}{c\tau_{N_2}}
\newcommand{\ctauH}{c\tau_{\tilde{\eta}^\pm}}
\newcommand{\mNtwo}{m_{N_2}}
\newcommand{\mNone}{m_{N_1}}
\newcommand{\mHpm}{m_{\tilde{\eta}^\pm}}
\newcommand{\meta}{m_{\tilde{\eta}^\pm}}
\newcommand{\DMopen}{\Delta M_{\rm open}}
\newcommand{\DMcomp}{\Delta M_{\rm comp}}
\newcommand{\MET}{\slashed{E}_T}
\newcommand{\fb}{\,\text{fb}}
\newcommand{\ifb}{\,\text{fb}^{-1}}
\newcommand{\GeV}{\,\text{GeV}}
\newcommand{\MeV}{\,\text{MeV}}
\newcommand{\ps}{\,\text{ps}}
\newcommand{\mm}{\,\text{mm}}
\newcommand{\RDV}{R_{\mathrm{DV}}}
\newcommand{\RDT}{R_{\mathrm{DT}}}
\newcommand{\dt}{\Delta t}

\journalname{Eur. Phys. J. C}

\begin{document}

\title{Nested Disappearing-Track plus Displaced-Vertex Topology
  at the HL-LHC: A Dual Long-Lived-Particle Signature
  of the Scotogenic Model}

\subtitle{}

\titlerunning{Nested DT$+$DV Dual-LLP Topology at the HL-LHC}

\author{Renjie Wang}

\authorrunning{R. Wang}

\institute{Institute of High Energy Physics,
  Chinese Academy of Sciences, Beijing 100049, China \\
  \email{rjwang@ihep.ac.cn}
}

\date{\today}

\abstract{%
The scotogenic model generates sub-eV neutrino masses radiatively
through small Yukawa couplings $\sim\mathcal{O}(10^{-3})$ that also
set the decay length of the heavy neutral fermion $\Ntwo$; in the
compressed regime a separate, much smaller Yukawa entry
$\sim\mathcal{O}(10^{-6})$ additionally renders the charged scalar
$\etapm$ long-lived.
We study the compressed-spectrum regime of its two-inert-doublet
variant ($\DMcomp = \mHpm - \mNtwo \simeq 200\MeV$), in which
small Yukawa entries render both the charged scalar $\etapm$ and
the heavy neutral fermion $\Ntwo$ long-lived and produce a nested
cascade at the High-Luminosity LHC~(HL-LHC): $\etapm$ leaves a
disappearing track~(DT), decays to an invisible $\Ntwo$, which
travels macroscopically before forming a displaced dilepton
vertex~(DV).
The two slow flights combine into a 4D time-of-flight delay
$\dt\sim200$--$800\ps$ resolvable by the CMS MIP Timing Detector,
while the DT-to-DV displacement vector inherits the parent track
direction, yielding a spatial back-pointing correlation with no
SM analogue.
These two handles are expected to suppress the background to
$B\sim10^{-5}$--$10^{-3}$ events, and the discovery remains
$5\sigma$-significant for any background up to $B\sim10$ events.
At the benchmark ($\mNtwo=150\GeV$ and proper decay lengths
$\ctauH=300\mm$, $\ctauN=189\mm$) the strategy yields
$N_s\approx1212$ signal events at $3000\ifb$ and establishes
sensitivity across $15\mm\lesssim\ctauH\lesssim300\mm$,
$10\mm\lesssim\ctauN\lesssim1000\mm$, a correlated region left
uncovered by any standalone DT or DV search.
Because the $\Ntwo$ decay length is controlled by the same Yukawa
couplings that generate sub-eV neutrino masses, mapping this
nested cascade turns the HL-LHC into a direct collider probe of
the interaction responsible for neutrino mass.
}

\maketitle

\section{Introduction}
\label{sec:intro}

The origin of neutrino mass remains one of the most compelling
open problems in particle physics.
The scotogenic model~\cite{Ma:2006km} provides an elegant
solution: the Standard Model~(SM) is extended by an inert scalar
doublet $\tilde{\eta}$ and right-handed Majorana singlets $N_k$,
stabilised by a $\mathbb{Z}_2$ symmetry, which simultaneously
forbids tree-level neutrino masses, generates them radiatively
at one loop, and renders the lightest odd-sector particle a
stable dark matter~(DM) candidate~\cite{Toma:2013zsa}.
It is one of the simplest and most predictive representatives of
the broad class of radiative neutrino mass
models~\cite{Cai:2017jrq}.
The two-inert-doublet variant (2IDM3N)~\cite{Ahriche:2023rtd}
further relaxes constraints from direct DM detection while
preserving the radiative mass-generation mechanism.

A central consequence of the scotogenic framework is that the
Yukawa coupling must satisfy $y \sim \mathcal{O}(10^{-3})$ to
reproduce observed sub-eV neutrino masses.
This tiny coupling renders the heavy neutral fermion $\Ntwo$
long-lived, with proper decay length
$\ctauN \sim \mathcal{O}(10\text{--}1000\mm)$
in the open-spectrum regime
($\DMopen \equiv \mNtwo - \mNone \sim 50\GeV$),
where $\etapm$ decays promptly and $\Ntwo$ is the sole LLP.

The present Letter explores a qualitatively different regime,
the \emph{compressed-spectrum} regime
($\DMcomp \equiv \mHpm - \mNtwo \simeq 200\MeV$), in which \emph{both}
$\etapm$ and $\Ntwo$ are simultaneously long-lived---the $\Ntwo$
lifetime governed by the $\mathcal{O}(10^{-3})$ neutrino-mass Yukawa
$y_1$ and the $\etapm$ lifetime by a much smaller entry
$y_2\sim\mathcal{O}(10^{-6})$.
This generates a \emph{nested topology}: $\etapm$ produces a
disappearing charged track~(DT) when it decays via the Yukawa
interaction as $\etapm \to \Ntwo + \ell^\pm_{\rm soft}$
($\ell = e,\mu$), with the soft lepton carrying momentum
$p_\ell \lesssim 0.2\GeV$, too soft to be reconstructed as an
isolated lepton and therefore not breaking the
disappearing-track signature.
$\Ntwo$ subsequently decays as
$\Ntwo \to \None + \ell^+\ell^-$
at a larger transverse radius, producing a 4D delayed
displaced vertex~(DV).

In the literature, scotogenic LLP searches in the compressed regime
have focused on $\etapm$ as the sole LLP, exploiting signatures
such as soft displaced leptons, ultra-compressed disappearing tracks,
or flavour-encoded timing~\cite{Alimena:2019zri}.
These analyses do not exploit the simultaneous macroscopic
lifetime of $\Ntwo$.
To our knowledge, the present work is the first study of the
nested DT$+$DV topology in the scotogenic benchmark,
producing a correlated spatial cascade unique in the SM;
correlated LLP topologies exist in other BSM
contexts~\cite{Alimena:2019zri} but none targets the
simultaneous DT$+$DV chain studied here.

This nested signature offers two decisive advantages:
\begin{enumerate}
  \item \textbf{Spatially correlated cascade.}
    The displacement vector from the DT endpoint to the DV
    position is required to align with the disappearing-track
    direction, consistent with an intermediate neutral $\Ntwo$
    emitted at the DT endpoint.
    This pointing condition has no SM analogue:
    no known SM process produces a high-$p_T$ charged track that
    disappears at transverse radius $\RDT \sim 10$--$300\mm$
    and is then followed, in the same flight direction, by a
    displaced dilepton vertex at transverse radius
    $\RDV > \RDT + 20\mm$.

  \item \textbf{Background-suppressing 4D timing.}
    The displaced leptons from $\Ntwo \to \None\,\ell^+\ell^-$
    carry a time-of-flight delay $\dt \sim 200$--$800\ps$
    measurable by the CMS MIP Timing
    Detector~(MTD)~\cite{CMSCollaboration:2667167}
    (and its ATLAS counterpart, the HGTD~\cite{ATLASCollaboration:HGTD};
    this paper focuses on the CMS MTD).
    A cut $\dt > 200\ps$, first proposed in
    Ref.~\cite{Liu:2018wte}, strongly suppresses the dominant
    $\gamma\to e^+e^-$ background and, combined with the
    spatial pointing requirement, suppresses instrumental
    backgrounds to $B\sim10^{-5}$--$10^{-3}$ events, with a discovery
    reach robust to $B\lesssim10$ events (Sect.~\ref{sec:analysis}).
\end{enumerate}

We demonstrate that the nested strategy covers a broad
two-dimensional parameter space in $(\ctauH, \ctauN)$ not
simultaneously targeted by any existing ATLAS or CMS
search~\cite{ATLAS:2022rme,CMS:2020atg,ATLAS:2023oti,CMS:2021kdm,%
ATLAS:2026hnb,CMS:2023mny,ATLAS:2024umc,CMS:2025qkk,%
ATLAS:2026riv,ATLAS:2026gcd,CMS:2024trg,ATLAS:2024vnc}.

The paper is organised as follows.
Section~\ref{sec:model} introduces the 2IDM3N model and derives
the dual LLP lifetimes.
Section~\ref{sec:signal} describes the nested signal topology and
production kinematics.
Section~\ref{sec:analysis} presents the analysis strategy,
cutflow, and discovery reach.
We conclude in Section~\ref{sec:conclusion}.

\section{Model and dual LLP lifetimes}
\label{sec:model}

\textbf{The 2IDM3N scotogenic model and benchmark.}
We work within the two-inert-doublet, three-right-handed-neutrino
extension of the scotogenic model~\cite{Ahriche:2023rtd},
hereafter 2IDM3N.
The particle content beyond the SM comprises two inert scalar
doublets $\eta_{1,2}$ and three Majorana fermion singlets
$N_{1,2,3}$, all odd under a $\mathbb{Z}_2$ symmetry.
The Yukawa interaction reads
\begin{equation}
  \mathcal{L} \supset
    y_{\alpha k}\,\overline{L_\alpha}\,\tilde{\eta}_1\,N_k
    + \tfrac{1}{2}\,M_k\,\overline{N_k^c}\,N_k
    + \text{h.c.},
\label{eq:yukawa}
\end{equation}
where $L_\alpha$ is the SM lepton doublet of generation $\alpha$
and $\tilde\eta_1 = i\sigma_2\eta_1^*$.
Neutrino masses are generated at one loop:
\begin{equation}
\begin{split}
(m_\nu)_{\alpha\beta}
    &= \sum_k \frac{y_{\alpha k}\, y_{\beta k}\, M_k}{16\pi^2} \\
    &\quad \times \left[
      \frac{m_R^2 \ln\left( \frac{m_R^2}{M_k^2} \right)}{m_R^2 - M_k^2}
      - \frac{m_I^2 \ln\left( \frac{m_I^2}{M_k^2} \right)}{m_I^2 - M_k^2}
    \right],
\label{eq:mnu}
\end{split}
\end{equation}
where $m_{R,I}$ are the real/imaginary neutral-scalar masses.
Sub-eV neutrino masses require
$y \sim \mathcal{O}(10^{-3})$ for
$M_k \sim \mathcal{O}(100\GeV)$.
The Yukawa coupling also mediates lepton-flavour-violating~(LFV)
processes such as $\mu \to e\gamma$, imposing complementary
constraints~\cite{Vicente:2014wga}.
For the small Yukawa entries relevant here, LFV bounds are
satisfied.

The compressed-spectrum scenario is defined by two mass
splittings, $\DMopen \equiv \mNtwo - \mNone = 50\GeV$ (open) and
$\DMcomp \equiv \mHpm - \mNtwo = 200\MeV$ (compressed),
with benchmark spectrum
$\mNtwo=150\GeV$, $\mNone=100\GeV$, $\mHpm=150.2\GeV$.
This point is representative of the compressed regime:
$\mNtwo=150\GeV$ is accessible to HL-LHC electroweak production,
$\DMcomp=200\MeV$ renders both $\etapm$ and $\Ntwo$ simultaneously
long-lived, and $\DMopen=50\GeV$ provides reconstructable
dilepton vertices.
Electroweak precision constraints are satisfied by setting
$m_{\tilde\eta^0}=m_{A_1^0}=150.2\GeV$ and decoupling the second doublet
at $m_{\eta_2}=5\,\text{TeV}$, which contributes to the
neutrino mass matrix through Eq.~(\ref{eq:mnu}) while playing
no role in collider phenomenology.
The collider signatures studied here are insensitive to the precise
value of $m_{\eta_2}$ as long as the second doublet is decoupled
($m_{\eta_2}\gtrsim1\,\text{TeV}$); the choice $5\,\text{TeV}$ is
representative, lies safely above the electroweak scale, and still
reproduces sub-eV neutrino masses through Eq.~(\ref{eq:mnu}) for
$\mathcal{O}(1)$ scalar couplings.

\textbf{\boldmath $\etapm$ lifetime (disappearing track).}
The Yukawa coupling of Eq.~(\ref{eq:yukawa}) is a matrix
$y_{\alpha k}$ in lepton flavour $\alpha$ and singlet generation
$k$.
Its $\Ntwo$ column, of effective magnitude
$y_2 \equiv (\sum_\alpha |y_{\alpha 2}|^2)^{1/2}$, governs the only
two-body decay of $\etapm$ available at this benchmark,
$\etapm \to \Ntwo + \ell^\pm$ ($\ell = e, \mu$);
the intra-doublet transition $\etapm \to \etaz\,\pi^\pm$ is closed
because $m_{\etapm} = m_{\etaz}$ here, so the charged-scalar
lifetime is fixed entirely by $y_2$.
The partial width in the nearly-degenerate limit
($\DMcomp \ll \mHpm$) is
\begin{equation}
  \Gamma(\etapm \to \ell^\pm \Ntwo)
  \approx \frac{y_2^2}{8\pi}
  \frac{(\DMcomp)^2}{\mHpm}\,N_\ell,
\label{eq:etawidth}
\end{equation}
where $N_\ell = 2$ sums the $e$ and $\mu$ modes
(the $\tau$ mode is kinematically closed: $m_\tau \gg \DMcomp$).
Because this is a renormalisable Yukawa decay rather than a
weak-scale--suppressed one, reaching a macroscopic disappearing
track $\ctauH \sim \mathcal{O}(10\text{--}300)\mm$ at
$\DMcomp = 200\MeV$ requires a small entry
$y_2 \sim \mathcal{O}(10^{-6})$, parametrically below the
couplings that set the neutral-fermion sector.
The $\Ntwo$ three-body width instead involves both the $\Ntwo$ and
the $\None$ columns and scales as
$\Gamma(\Ntwo \to \None\,\ell^+\ell^-) \propto y_2^2\,y_1^2$ in the
Yukawa couplings, with $y_1 \equiv (\sum_\alpha |y_{\alpha 1}|^2)^{1/2}$;
the mass dependence is evaluated numerically (see below), so that
\begin{equation}
  \ctauH \propto \frac{1}{y_2^2\,(\DMcomp)^2},
  \qquad
  \ctauN \propto \frac{1}{y_2^2\,y_1^2}\Big|_{\rm fixed\ masses}.
\label{eq:lifetimes}
\end{equation}
For fixed masses $\ctauN$ scales as $1/(y_1^2 y_2^2)$, while the full
mass dependence is evaluated numerically with \textsc{MadGraph5}
because the off-shell $\etapm$ propagator is near threshold for
$\DMcomp = 200\MeV$: with $\mHpm-\mNtwo = 200\MeV$ the
virtual-$\etapm$ momentum transfer approaches $m_{\etapm}^2$, so the
width is enhanced relative to the naive $1/\meta^4$ contact estimate.
The two proper decay lengths are thus controlled by
\emph{independent} entries of the Yukawa matrix --- $y_2$ alone
for $\ctauH$, and $y_1$ at fixed $y_2$ for $\ctauN$ --- and can be
dialled independently.  We accordingly treat $\ctauH$ and $\ctauN$
as the two independent long-lived-particle parameters of the
problem, realised by the underlying $(y_2,\,y_1,\,\DMcomp)$, as is
standard in LLP phenomenology, and map the sensitivity over the
full $(\ctauH, \ctauN)$ plane rather than committing to a single
ultraviolet-complete point.
The Yukawa entries that, together with a small neutral-scalar
splitting, reproduce sub-eV neutrino masses through
Eq.~(\ref{eq:mnu}) lie at the $\mathcal{O}(10^{-3})$ level; since
the $\Ntwo$ decay length is set by these same couplings, the
displaced-vertex side of the cascade directly probes the
neutrino-mass-generating sector, while the charged-scalar track
length is an independent compressed-spectrum handle governed by
the smaller $y_2$.

The soft lepton from $\etapm \to \ell^\pm \Ntwo$ carries
momentum $p_\ell \simeq 0.17$--$0.20\GeV$ in the $\etapm$ rest
frame (the two-body K\"all\'en value for $\DMcomp = 200\MeV$,
the upper edge for $\ell = e$ and the lower for $\ell = \mu$);
the mild $\etapm$ boost ($\gamma\simeq1.1$--$1.5$) keeps the
lab-frame momentum below ${\sim}0.5\GeV$.
A lepton this soft is not reconstructed as an isolated object that
could veto the disappearing track or extend the $\etapm$ stub.
A muon with $p\lesssim0.5\GeV$ falls well below the
$p>2.5\GeV$ (and $p_T>0.5\GeV$) requirement for a tracker track to
be extrapolated to the muon system and identified as a
muon~\cite{CMS:2018rym}, and in any case ranges out in the
calorimeters and return yoke before reaching the muon chambers;
an electron of comparable momentum deposits only a small
electromagnetic cluster and lies far below the $p_T\gtrsim5\GeV$
isolated-lepton thresholds applied both in this analysis and in
the displaced-lepton searches discussed in
Sect.~\ref{sec:analysis}.
This soft lepton plays exactly the role of the soft pion in
chargino disappearing-track searches
($\chi^\pm\to\chi^0\pi^\pm$, $p_\pi\sim0.1$--$0.2\GeV$), where the
soft daughter is likewise unreconstructed and the high-$p_T$ parent
track simply disappears~\cite{ATLAS:2022rme,CMS:2020atg}.
The lepton is moreover emitted nearly collinearly with the parent
($\delta\theta \lesssim 3\,\text{mrad}$, see
Sect.~\ref{sec:signal}), so the residual soft-lepton reconstruction
efficiency and any associated disappearing-track veto inefficiency
are small; their precise size would be quantified by a full detector
simulation at $\langle\mu\rangle = 200$.
We scan $\ctauH \in \{1,\,5,\,15,\,30,\,50,\,150,\,300\}\mm$
to cover the range accessible to the CMS barrel pixel detector
($R_{\rm DT} \geq 160\mm$, see Sect.~\ref{sec:analysis})
through the outer tracker
($R \gtrsim 200\mm$)~\cite{CMS:TrackerTDR}.

\textbf{\boldmath $N_2$ lifetime (displaced vertex).}
Because $\Ntwo$ is $\mathbb{Z}_2$-odd, its only available decay
channel is the three-body process
$\Ntwo \to \None + \ell^+ + \ell^-$ ($\ell = e,\,\mu$),
mediated by an off-shell $\etapm$.
Tau leptons are not considered: their displaced hadronic or
semi-leptonic decay products require a substantially more complex
vertex reconstruction algorithm, and their inclusion is left for
future work.
The partial width scales as
$\Gamma(\Ntwo \to \None\,\ell^+\ell^-) \propto y_2^2\,y_1^2$ in the
Yukawa couplings, with the remaining mass dependence---non-trivial
because the off-shell $\etapm$ propagator is near threshold for
$\DMcomp=200\MeV$---evaluated numerically; from this
$\ctauN=\hbar c/\Gamma$.
As a representative point we adopt $\ctauN \approx 189\mm$,
corresponding to a three-body width
$\Gamma = 1.045 \times 10^{-15}\,\text{GeV}$
(verified by a dedicated \textsc{MadGraph5}~\cite{Alwall:2014hca}
calculation); Yukawa entries of the neutrino-mass scale
($y_1 \sim \mathcal{O}(10^{-3})$, at the small $y_2$ fixed by the
disappearing-track requirement) yield $\Ntwo$ decay lengths
across the probed range.
We scan eight $\ctauN$ values from 10 to 1000\,mm
(10, 30, 60, 120, 189, 400, 700, and 1000\,mm),
with 189\,mm taken as the representative point for the cutflow.

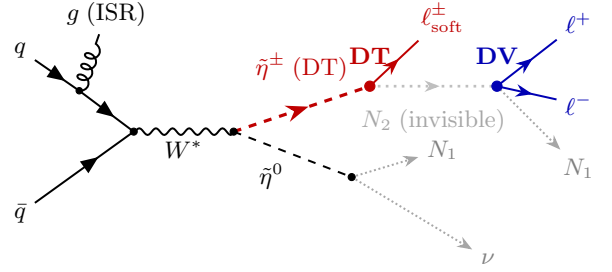
\begin{figure}[t]
\centering
\begin{tikzpicture}[scale=0.6]
\begin{feynman}[every edge={line width=0.75pt}]

  \vertex (i1)   at (-5.5,  1.8)  {$q$};
  \vertex (i2)   at (-5.5, -1.8)  {$\bar{q}$};
  \vertex (visr) at (-4.2,  0.9);
  \vertex (isr)  at (-3.6,  2.6)  {$g\;(\text{ISR})$};
  \vertex (vW)   at (-3.0,  0.0);

  \vertex (vP)   at (-0.8,  0.0);

  \vertex (vD0)  at ( 1.8, -1.0);
  \vertex (fN1c) at ( 3.8, -0.4) {\textcolor{gray}{$N_1$}};
  \vertex (fnu)  at ( 4.8, -2.8) {\textcolor{gray}{$\nu$}};

  \vertex (vDT)  at ( 2.2,  1.0);
  \vertex (fsoft)at ( 3.8,  2.5)
      {\textcolor{red!75!black}{$\ell^\pm_{\rm soft}$}};

  \vertex (vDV)  at ( 5.0,  1.0);
  \vertex (felp) at ( 6.8,  2.4) {\textcolor{blue!70!black}{$\ell^+$}};
  \vertex (felm) at ( 6.8,  0.6) {\textcolor{blue!70!black}{$\ell^-$}};
  \vertex (fN1a) at ( 6.8, -0.8) {\textcolor{gray}{$N_1$}};

  \diagram*{
    (i1)   -- [fermion]      (visr),
    (visr) -- [fermion]      (vW),
    (vW)   -- [anti fermion] (i2),
    (visr) -- [gluon]        (isr),
    (vW)   -- [boson, edge label'=$W^*$]           (vP),
    (vP)   -- [scalar, edge label'=$\tilde\eta^0$] (vD0),
  };

  \draw[gray!70, densely dotted, line width=0.75pt, -{Stealth[length=2mm]}]
      (vD0) -- (fN1c);
  \draw[gray!70, densely dotted, line width=0.75pt, -{Stealth[length=2mm]}]
      (vD0) -- (fnu);

  \draw[red!75!black, very thick, dashed,
        postaction={decorate, decoration={markings,
          mark=at position 0.55 with {\arrow{Stealth}}}}]
      (vP) -- (vDT)
      node[midway, above=7pt, font=\small, red!75!black]
      {$\tilde\eta^\pm$\;(DT)};

  \draw[red!75!black, line width=0.75pt,
        postaction={decorate, decoration={markings,
          mark=at position 0.6 with {\arrow{Stealth}}}}]
      (vDT) -- (fsoft);

  \draw[gray!50, line width=1.0pt, dotted,
        postaction={decorate, decoration={markings,
          mark=at position 0.55 with {\arrow{Stealth[length=2mm]}}}}]
      (vDT) -- (vDV)
      node[midway, below=5pt, font=\small, gray!70]
      {$N_2$\;(invisible)};

  \draw[blue!70!black, line width=0.75pt,
        postaction={decorate, decoration={markings,
          mark=at position 0.6 with {\arrow{Stealth}}}}]
      (vDV) -- (felp);
  \draw[blue!70!black, line width=0.75pt,
        postaction={decorate, decoration={markings,
          mark=at position 0.6 with {\arrow{Stealth}}}}]
      (vDV) -- (felm);
  \draw[gray!70, densely dotted, line width=0.75pt, -{Stealth[length=2mm]}]
      (vDV) -- (fN1a);

  \filldraw[black]           (visr) circle (2pt);
  \filldraw[black]           (vW)   circle (2.2pt);
  \filldraw[black]           (vP)   circle (2.2pt);
  \filldraw[black]           (vD0)  circle (2.2pt);
  \filldraw[red!75!black]    (vDT)  circle (3.2pt);
  \filldraw[blue!70!black]   (vDV)  circle (3.2pt);

  \node[above=5pt, red!75!black, font=\small\bfseries]  at (vDT) {DT};
  \node[above=5pt, blue!70!black, font=\small\bfseries] at (vDV) {DV};

\end{feynman}
\end{tikzpicture}
\caption{Representative signal topology for
  $pp\to W^*\to\tilde\eta^\pm\tilde\eta^0+\text{ISR}$.
  The charged scalar $\tilde\eta^\pm$ (red dashed,
  $\ctauH \sim \mathcal{O}(10\text{--}300)\mm$)
  propagates macroscopically before decaying at the disappearing-track
  endpoint~(DT, red circle) into a soft lepton $\ell^\pm_{\rm soft}$
  (too soft to reconstruct as an isolated lepton) and a neutral $N_2$.
  The $N_2$ travels invisibly (grey dotted,
  $\ctauN \sim \mathcal{O}(10\text{--}1000)\mm$)
  to a second macroscopic displacement, decaying at the displaced
  vertex~(DV, blue circle) as
  $N_2\to N_1\,\ell^+\ell^-$; $N_1$ escapes as missing energy.
  The neutral $\tilde\eta^0$ decays invisibly into $\nu N_1$.}
\label{fig:feynman}
\end{figure}

Figure~\ref{fig:feynman} illustrates the nested signal topology
for the dominant production channel
$pp\to W^*\to\tilde\eta^\pm\tilde\eta^0 + \text{ISR}$.

\section{Signal topology and production}
\label{sec:signal}

\textbf{Production channels.}
Signal events are produced via electroweak pair production:
\begin{align}
  pp &\to \etapm\,\etaz,\quad
  pp \to \etapm\,A_1^0,\quad
  pp \to \tilde{\eta}^+\tilde{\eta}^-,
\label{eq:production}
\end{align}
mediated by $s$-channel $W^\pm$, $Z$, and $\gamma$ exchange,
where $\etaz$ and $A_1^0$ are the CP-even and CP-odd neutral
scalar components of $\eta_1$, respectively.
The two $\etapm$-associated channels ($\etapm\etaz$ and
$\etapm A_1^0$) contribute roughly half the total cross section
via $W^\pm$-mediated production and are included in the signal.
The fiducial LO cross section, evaluated with
\textsc{MadGraph5}~\cite{Alwall:2014hca} using
PDF4LHC21\_40~\cite{PDF4LHC21:2022mnw}
after generator-level cuts slightly looser than the
Path~A trigger requirements
($p_T^j > 100\GeV$, $|\eta_j| < 2.8$, $\MET^{\rm gen} > 140\GeV$),
is $\sigma_{\rm fid}^{\rm LO} = 27.39\,^{+13.6\%}_{-11.3\%}\,(\text{scale})\,^{+2.5\%}_{-1.7\%}\,(\text{PDF})\fb$,
where the scale (PDF) uncertainty is the seven-point $\mu_R,\mu_F$
envelope (PDF4LHC21 replica spread) returned by \textsc{MadGraph5}.
Applying a uniform NLO $K$-factor
$K = 1.25$~\cite{Fuks:2012qx,Fuks:2013lya} gives
$\sigma_{\rm fid}^{\rm NLO} = 34.24\fb$, to which the same relative
scale and PDF uncertainties apply (the LO scale band is conservative,
as it shrinks at NLO).
The gauge quantum numbers of $\tilde{\eta}$ match those
of the inert doublet model~\cite{Belyaev:2018ext}.

\textbf{Nested 4D signal topology.}
The complete detector event unfolds in four spacetime dimensions:

\begin{enumerate}
  \item \textbf{$t=0$, Primary vertex~(PV).}
    The electroweak processes of Eq.~(\ref{eq:production})
    produce a $\etapm$ together with a neutral partner
    ($\etaz$, $A_1^0$, or $\tilde{\eta}^\mp$).
    Because $\DMcomp = 200\MeV$, no hard lepton is available
    from the $\etapm$ decay; instead, an initial-state
    radiation~(ISR) jet provides the Level-1 trigger seed
    (Sect.~\ref{sec:analysis}).

  \item \textbf{$t \approx 0$--$1\,\text{ns}$,
    Disappearing track~(DT).}
    Each $\etapm$ propagates through the inner tracker as a
    high-$p_T$ charged track, decaying at transverse radius
    $\RDT \sim \ctauH\langle\gamma\beta\rangle_\pm\sin\theta$ as
    $\etapm \to \Ntwo + \ell^\pm_{\rm soft}$.
    The soft lepton ($p_\ell \lesssim 0.2\GeV$) is too soft to
    form an isolated reconstructed lepton and the $\etapm$ track
    abruptly disappears.

  \item \textbf{$t = t_{\rm DV}$, Displaced vertex~(DV).}
    The $\Ntwo$, electrically neutral and $\mathbb{Z}_2$-odd,
    propagates invisibly from the DT endpoint.
    At $\RDV > \RDT + 20\mm$, it decays as
    $\Ntwo \to \None + \ell^+ + \ell^-$, producing two
    reconstructed leptons at a displaced point with
    $\RDV \in [180,\,700]\mm$, and $\None$ escapes as $\MET$.

\end{enumerate}

\textbf{$N_2$ kinematics and dilepton separation.}
Because $\DMcomp = 200\MeV \ll \mNtwo$, the $\Ntwo$ receives
negligible recoil from the $\etapm \to \Ntwo + \ell^\pm_{\rm soft}$ decay
and inherits the full $\etapm$ lab-frame momentum.
For $\mHpm = 150.2\GeV$ and typical $p_T^{\etapm} \sim 75\GeV$,
the $\Ntwo$ boost is
$\gamma_{\Ntwo} \approx \gamma_{\etapm} \approx
\sqrt{1 + (p_T/m)^2} \approx 1.1$--$1.5$
($\beta\gamma \sim 0.5$--$1.1$).

In the $\Ntwo$ rest frame, the three-body decay
$\Ntwo \to \None + \ell^+ + \ell^-$ distributes the available
kinetic energy $Q = \DMopen = 50\GeV$ among the three final-state
particles.
Phase space favours configurations where $N_1$ is emitted
back-to-back with the dilepton system, giving typical rest-frame
lepton momenta $p_\ell^* \sim Q/3 \approx 17\GeV$.
The rest-frame opening angle between $\ell^+$ and $\ell^-$
spans the full kinematic range; for the dominant phase-space
region the two leptons are not collinear
($\theta^*_{\ell\ell} \sim 60^\circ$--$150^\circ$).

After boosting to the lab frame, the moderate $\Ntwo$ boost
($\gamma \sim 1$--$2$) preserves a large opening angle:
the lab-frame angular separation between the two leptons is
$\Delta\theta_{\ell\ell} \sim 1/\gamma \sim 0.5$--$1\,\text{rad}$,
ensuring they strike the MTD barrel at well-separated azimuthal
positions and produce two geometrically distinct timing hits.
Since the leptons emerge from a displaced vertex at
$\RDV \in [180,\,700]\mm$, their impact parameters satisfy
$d_0 \gtrsim \RDV \sin\alpha \gg 2\mm$ for any non-radial
emission angle $\alpha$, comfortably passing the Path~B
displaced-muon trigger requirement ($|d_0| > 2\mm$).
The large opening angle also means no coincidence window
$|\dt_1-\dt_2|$ need be imposed: the geometric TOF spread
between the two leptons (median ${\sim}200\ps$) is
unrelated to the signal delay $\dt$.

\begin{enumerate}
\setcounter{enumi}{3}

  \item \textbf{$t = t_{\rm MTD}$, Delayed timing.}
    The displaced leptons reach the MTD barrel
    ($R = 1170\mm$) with an arrival-time delay comprising
    \emph{two} contributions, a defining feature of the
    nested topology:
    \begin{equation}
    \begin{split}
      \dt \approx\;
        &\frac{\RDT}{c}\!\left(\frac{1}{\beta_{\etapm}}-1\right) \\
        &+\frac{\RDV - \RDT}{c}\!\left(\frac{1}{\beta_{\Ntwo}}-1\right)
        + \mathcal{O}(\sigma_t),
    \end{split}
    \label{eq:deltat}
    \end{equation}
    where $\sigma_t = 30\ps$~\cite{CMSCollaboration:2667167}.
    The first term is the delay from the slow $\etapm$
    ($\beta_{\etapm} < 1$, PV$\to$DT); the second from
    the slow $\Ntwo$ (DT$\to$DV).
    For $\mHpm \approx \mNtwo = 150\GeV$,
    $\beta_{\etapm} \sim \beta_{\Ntwo} \sim 0.7$,
    giving $\dt \sim 200$--$800\ps$, shifted to larger
    values than in single-LLP scenarios.
\end{enumerate}

\textbf{Spatial pointing condition.}
The displacement vector from the DT endpoint to the DV must align
with the \emph{disappearing-track direction}
($\Delta\phi_{\rm DT\to DV},\Delta\theta_{\rm DT\to DV}<0.1\,\text{rad}$).
This is the actual discriminant: since the $\Ntwo$ is emitted at the
DT endpoint, the DT$\to$DV vector is by construction the $\Ntwo$ flight
direction, so the cut tests whether that direction is inherited from the
parent $\etapm$ track --- a correlation absent in random DT$+$DV pairings.

\emph{$\Ntwo$ collinearity with $\etapm$:}
The soft lepton carries $p_\ell\lesssim0.2\GeV\ll p_{\etapm}$,
so $\delta\theta\sim p_\ell/p_{\etapm}\lesssim3\,\text{mrad}$,
far below the cut; in signal simulation the reconstructed DT-to-DV
back-pointing angle peaks at ${\approx}2.5\,\text{mrad}$, so the
pointing requirement is essentially fully efficient.

\emph{Pointing observable and $N_1$ impact:}
If one attempted a \emph{momentum-based} pointing variable, the
$\Ntwo$ direction would have to be approximated by the dilepton sum
$\vec{p}_\Sigma\equiv\vec{p}_{\ell^+}+\vec{p}_{\ell^-}$, since $\None$
escapes as $\MET$.
In the $\Ntwo$ rest frame, $\None$ carries
$p_{\None}^*\sim\DMopen/3\approx17\GeV$ while the dilepton
system carries ${\sim}2p_{\None}^*\approx34\GeV$, giving a
lab-frame misalignment
\begin{equation}
  \delta\theta_{\rm point} \sim
  \frac{p_{\None}^*}{|\vec{p}_\Sigma^{\,*}|} \cdot
  \frac{1}{\gamma_{\Ntwo}}
  \approx \frac{1}{2} \times \frac{1}{1.3}
  \approx 0.38\,\text{rad}.
\label{eq:pointing_approx}
\end{equation}
The actual analysis avoids this limitation altogether by using the
\emph{geometric} displacement vector
$\vec{r}_{\rm DT\to DV}$ — the three-dimensional vector
from the reconstructed DT endpoint to the reconstructed
DV position — which is a purely spatial quantity.
It is \emph{not} the dilepton momentum sum $\vec{p}_\Sigma$,
and it does not involve any MET reconstruction.
Consequently, the MET from the $\Ntwo\to\None$ decay,
the neutral partner ($\etaz$, $A_1^0$) decay,
or any other invisible particle in the event
has \emph{no effect} on the pointing cut:
the geometric alignment between the DT endpoint
and the DV position is entirely determined by the
flight path of the invisible $\Ntwo$, regardless
of what else is produced in the event.
The $\delta\theta_{\rm point}\approx0.38\,\text{rad}$
estimated above is relevant only for a \emph{momentum-based}
pointing cut, which a realistic analysis might use
as an alternative or cross-check.
In the analysis (Sect.~\ref{sec:analysis}) the pointing cut is
evaluated on the \emph{reconstructed} DV position relative to the
disappearing-track direction, not on the dilepton-sum approximation;
the only truth-level input retained anywhere in the selection is the
DT endpoint radius $\RDT$, since \textsc{Delphes} does not reconstruct
the pixel stub. Residual efficiency losses from MTD acceptance and the
$N_1$ fraction would be quantified by a full pileup-overlay simulation.
\emph{Choice of pointing window:} the $0.1\,\text{rad}$ requirement is
deliberately ${\sim}30$ times looser than the ${<}3\,\text{mrad}$
physical collinearity of the signal. It is set not by the intrinsic
alignment but by the back-pointing resolution of the reconstructed DT
endpoint and DV position; a dedicated simulation of the benchmark
yields a reconstructed back-pointing angle peaking at
${\approx}2.5\,\text{mrad}$, well within the cut. We adopt the
conservative $0.1\,\text{rad}$ window to absorb position-resolution
tails at $\langle\mu\rangle=200$, and---crucially---the background
estimate below applies the \emph{same} $0.1\,\text{rad}$ window
($P_{\rm point}\approx(0.1)^2/\pi$), so signal and background are
evaluated with an identical, conservative pointing requirement.
The pointing condition has no SM analogue and provides
strong, though not absolute, background suppression.

\begin{figure}[ht]
  \centering
  \includegraphics[width=0.9\columnwidth]{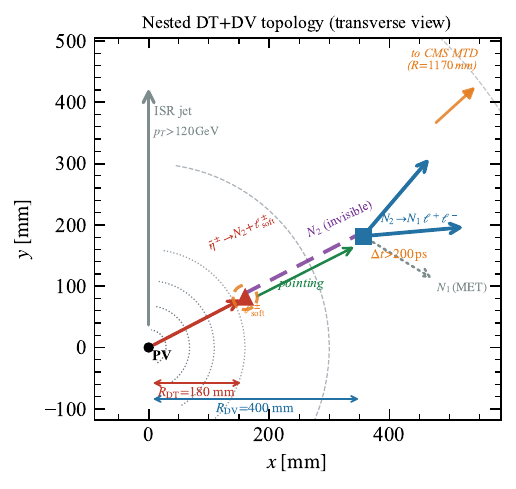}
  \caption{Nested DT$+$DV topology in the transverse plane
    (benchmark: $\RDT = 180\mm$, $\RDV = 400\mm$).
    Colour code: red = $\etapm$ track and DT endpoint (triangle);
    orange dashed curl = soft lepton $\ell^\pm_{\rm soft}$
    (too soft to reconstruct, $p_\ell \lesssim 0.2\GeV$);
    purple dashed = invisible $\Ntwo$ flight;
    blue = DV (square) and displaced lepton arrows ($\ell^+$, $\ell^-$);
    grey dotted = $\None$ escaping as missing energy (MET);
    green = pointing condition (DT endpoint $\to$ DV);
    grey solid = ISR jet;
    orange outward arrow = direction to the CMS MTD
    ($R = 1170\mm$, beyond the frame), reached by the
    displaced leptons where the timing $\dt > 200\ps$ is recorded.
    The $\None$ arrow illustrates that only the visible dilepton
    sum $\vec{p}_{\ell^+}+\vec{p}_{\ell^-}$ is available to
    reconstruct the $\Ntwo$ direction; see text.}
  \label{fig:topology}
\end{figure}

Figure~\ref{fig:topology} illustrates the complete nested topology
for the benchmark scenario ($\RDT = 180\mm$, $\RDV = 400\mm$).
The $\etapm$ (red arrow) propagates from the PV to the DT
endpoint (red triangle), where it decays into $\Ntwo$ plus an
undetectable soft lepton $\ell^\pm_{\rm soft}$ (dashed curl).
The electrically neutral $\Ntwo$ (purple dashed segment) travels
invisibly to the DV (blue square), where it decays as
$\Ntwo \to \None\,\ell^+\ell^-$; the two displaced leptons (blue
arrows) arrive at the CMS MTD with delay $\dt > 200\ps$, while
$\None$ escapes as missing energy (grey dotted arrow).
The green arrow shows the spatial pointing condition
(DT endpoint $\to$ DV), enforced by the small soft-lepton momentum
($p_\ell \lesssim 0.2\GeV$; see Sect.~\ref{sec:signal}).
An ISR jet (grey arrow) provides the Level-1 trigger seed.

\section{Analysis strategy and results}
\label{sec:analysis}

\textbf{Simulation and triggers.}
Signal events are generated with
\textsc{MadGraph5\_aMC@NLO} v3.5.1~\cite{Alwall:2014hca}
at $\sqrt{s} = 14\,\text{TeV}$
using the 2IDM3N UFO model~\cite{Ahriche:2023rtd}.
The three electroweak channels of Eq.~(\ref{eq:production}),
$pp \to \etapm\etaz$, $\etapm\Aone$, and $\tilde\eta^+\tilde\eta^-$,
are generated with an associated initial-state radiation jet, imposing generator-level phase-space
cuts ($p_T^j > 100\GeV$, $|\eta_j| < 2.8$,
$\MET^{\rm gen} > 140\GeV$) that are slightly looser than the
Path~A trigger requirements (discussed in Sect.~\ref{sec:analysis})
and suppress the soft phase-space region.
Parton showering and hadronisation are performed with
\textsc{Pythia~8.311}~\cite{Bierlich:2022pfr}.

Both $\etapm$ and $\Ntwo$ are treated as long-lived particles
with macroscopic proper decay lengths.
For each benchmark point in the lifetime scan, the proper
decay lengths $\ctauH$ and $\ctauN$ are set directly in
\textsc{Pythia~8} without regenerating the hard process,
since the electroweak production cross section is independent
of the Yukawa coupling.
$\Ntwo$ is declared as a Majorana fermion
(self-conjugate under charge conjugation)
with equal branching fractions to
$e^+e^-$ and $\mu^+\mu^-$ final states, as required by
the Yukawa structure of Eq.~(\ref{eq:yukawa})~\cite{Toma:2013zsa}.
All decay-length limits inside \textsc{Pythia~8} are disabled
to allow decays anywhere in the detector volume.
Detector simulation uses
\textsc{Delphes~3}~\cite{deFavereau:2013fsa}
with the CMS Phase-II card,
$\sigma_t = 30\ps$~\cite{CMSCollaboration:2667167}.

A note on DT reconstruction: \textsc{Delphes~3} does not
simulate hardware-level pixel-hit counting.
The CMS Phase-2 barrel pixel detector has four layers at radii
$\approx 29$, $68$, $109$, and $160\mm$~\cite{CMS:TrackerTDR};
a track decaying at $\RDT < 160\mm$ passes through fewer than
four barrel pixel layers and cannot reliably satisfy the
$\geq 3$--$4$ hit requirements used in real DT
searches~\cite{ATLAS:2022rme,CMS:2020atg,ATLAS:2026hnb,CMS:2023mny}.
While a recent ATLAS search~\cite{ATLAS:2026hnb} explicitly
reconstructs short tracks with as few as three innermost-layer hits,
reaching proper decay lengths down to ${\sim}3\mm$, we conservatively
require $\geq 4$ barrel pixel-layer crossings.
We therefore adopt the lower bound
$\RDT \geq 160\mm$, counting a track as a DT candidate only if
$\RDT \in [160,\,300]\mm$.
This ensures $\geq 4$ barrel pixel-layer crossings and maximises
confidence in the reconstructability of the stub.
Unit DT reconstruction efficiency is assumed within this window,
a standard approximation in phenomenological
studies~\cite{Alimena:2019zri,Cerri:2018rkm};
a dedicated experimental study with full pileup simulation
($\langle\mu\rangle = 200$) is needed to quantify residual
efficiency corrections at HL-LHC conditions.

\textbf{Reconstruction-level analysis.}
With the single exception of the DT endpoint radius above, every
observable that enters the selection is reconstruction-level rather
than truth-level, directly addressing the concern that idealised,
truth-matched quantities could drive the sensitivity.
The 4D-timing reference time $t_0$ is taken from the reconstructed
hard-scatter primary vertex, obtained from a four-dimensional (space
and time) fit of the timed tracks, rather than from a truth interaction
time. The per-vertex arrival-time delay $\dt$ and the late-track
fraction are then formed from the MTD times of the displaced tracks
measured relative to this reconstructed $t_0$. The displaced vertex is
built by clustering the production points of reconstructed displaced
tracks in three dimensions, and is required to contain a reconstructed
same-flavour $e^+e^-$ or $\mu^+\mu^-$ pair with $p_T > 5\GeV$ and
$|d_0| > 2\mm$ whose displacement vector back-points to the DT endpoint;
no truth lepton or truth vertex is used.
Pile-up is suppressed by the MTD time itself: a track is discarded only
if it arrives \emph{earlier} than $t_0$ beyond the timing resolution,
since an early track can only originate from an out-of-time pile-up
vertex, whereas the prompt and genuinely-late daughters of a long-lived
$\Ntwo$ are retained. This time-asymmetric cleaning is therefore
signal-safe and, unlike a truth pile-up flag, remains applicable at the
HL-LHC pile-up of $\langle\mu\rangle = 200$. The missing transverse
momentum and the trigger objects (the Path~A ISR jet and the Path~B
displaced muon) are likewise reconstructed quantities. The pointing,
timing, and vertex requirements that provide the dominant background
rejection are thus all evaluated on detector-level observables rather
than on truth matching.

We simulate $5\times10^4$ events per benchmark point on a
$7\times8$ grid spanning
$\ctauH \in [1,\,300]\mm$ (7 values: 1, 5, 15, 30, 50, 150, 300\,mm) and
$\ctauN \in [10,\,1000]\mm$ (8 values),
totalling 56 benchmark points (${\sim}2.8\times10^6$ events).

Since $\DMcomp = 200\MeV$ renders the soft lepton from $\etapm$
undetectable and $\Ntwo$ is neutral, no hard charged object is
available at the primary vertex.
We employ a logical OR of two complementary Level-1 trigger paths.

\noindent\textbf{Path~A: ISR jet plus $\MET$.}
Requires $p_T^j > 120\GeV$, $|\eta_j| < 2.5$, and $\MET > 150\GeV$,
corresponding to CMS L1 path
\mbox{\texttt{L1\_SingleJet120\_MET150}} with efficiency
$\varepsilon_A \approx 0.85$, taken from the measured L1 trigger
performance in Ref.~\cite{CMS:2020cmk}.

\noindent\textbf{Path~B: Phase-II displaced muon trigger.}
For the $\Ntwo \to \mu^+\mu^-$ sub-sample (50\% BR), a
displaced muon with $p_T^\mu > 15\GeV$, $|\eta^\mu| < 2.4$,
and $|d_0| > 2\mm$
provides a trigger seed at the CMS Phase-II L1 Track
Trigger~\cite{CMS:2020cmk_L1TDR} with efficiency
$\varepsilon_B \approx 0.70$, as projected in the Phase-2
L1 Trigger TDR~\cite{CMS:2020cmk_L1TDR}.
The combined OR efficiency, for events containing the trigger
objects, is
$\varepsilon_{\rm trig} = 1-(1-\varepsilon_A)(1-\varepsilon_B^{\rm eff})
\approx 0.90$,
where $\varepsilon_B^{\rm eff} = 0.50\times\varepsilon_B$.
In the cutflow of Table~\ref{tab:cutflow} the trigger is applied
\emph{after} the disappearing-track requirement, where it retains
$5629/10681\approx53\%$ of DT-tagged candidates, since only a
fraction of DT events also carry an ISR jet or a displaced muon
above the Level-1 thresholds.

\textbf{Signal region and cutflow.} %
We define the signal region SR-DT4DV by the following
sequential requirements:
\begin{enumerate}
  \item \textbf{Trigger.}
    Path~A or Path~B (Sect.~\ref{sec:analysis}).
  \item \textbf{Disappearing track~(DT) candidate.}
    $\geq 1$ track with $p_T > 50\GeV$, $|\eta| < 2.1$,
    $160\mm \leq L_{\rm track} \leq 300\mm$,
    no ECAL/HCAL deposit beyond the endpoint,
    and $\Delta R > 0.4$ from any jet.
  \item \textbf{Displaced dilepton vertex.}
    A displaced vertex clustered from $\geq 2$ displaced tracks,
    at least one matched to a reconstructed lepton ($e$ or $\mu$;
    both where reconstruction succeeds, forming a same-flavour
    pair),
    with $\RDV \in [180,700]\mm$, matched-lepton $p_T > 5\GeV$,
    $\RDV > \RDT + 20\mm$, and $m_{\ell\ell} < 48\GeV$ applied
    only when a same-flavour pair is reconstructed.
    The 20\,mm separation is required to ensure
    that the DV lepton tracks do not share pixel hits
    with the DT stub and that the DV is not
    misidentified as a nuclear interaction at the
    DT endpoint; it is confirmed to have negligible
    efficiency loss for signal ($\RDV \gg \RDT$ for
    $\ctauN \gg \ctauH$, as at the benchmark).
  \item \textbf{4D timing cut.}
    The displaced vertex carries a mean arrival-time delay
    $\dt > 200\ps$, formed from its MTD-timed displaced tracks
    relative to the reconstructed primary-vertex time $t_0$;
    this strongly suppresses
    $\gamma\!\to\!e^+e^-$~\cite{Liu:2018wte}
    ($\beta_\gamma=1 \Rightarrow \dt\approx 0$).
    Because a single per-vertex delay is used, no coincidence
    window $|\dt_1-\dt_2|$ between the two leptons is imposed
    (see kinematics discussion above).
  \item \textbf{Spatial pointing.}
    $\Delta\phi_{\rm DT\to DV},\,\Delta\theta_{\rm DT\to DV}<0.1\,\text{rad}$
    (see Sect.~\ref{sec:signal}).
  \item \textbf{Missing transverse energy.}
    $\MET > 50\GeV$.
    This offline threshold is deliberately kept low to maximise
    signal acceptance from the displaced-muon trigger (Path~B),
    which recovers events with soft ISR jets.
    For events relying on Path~A alone, the effective kinematic
    bound is governed by the higher L1 requirement
    ($\MET^{\rm L1} > 150\GeV$), so this offline cut is not
    the limiting constraint in that sub-sample.
\end{enumerate}

\begin{figure}[ht]
  \centering
  \includegraphics[width=0.9\columnwidth]{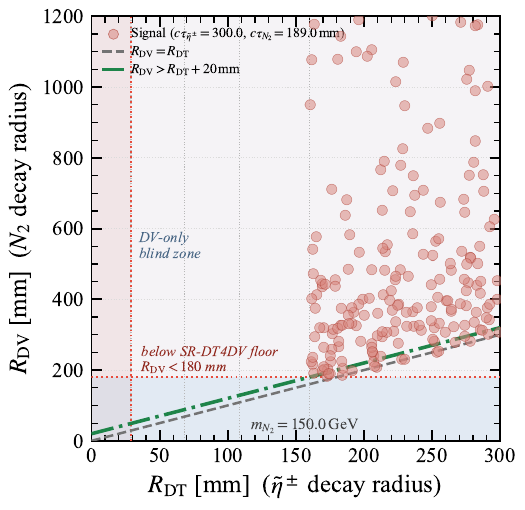}
  \caption{Signal distribution in the $(\RDT,\,\RDV)$ plane
    for the benchmark ($\ctauH = 300\mm$, $\ctauN = 189\mm$,
    $50\,000$ events).
    Shaded: the DV-only blind zone (left) and the region below the
    effective displaced-vertex floor $\RDV<180\mm$; this floor is not an
    independent cut but follows from $\RDT\ge160\mm$ together with the
    nested requirement $\RDV>\RDT+20\mm$ (green line).
    Grey dashed: $\RDV = \RDT$ diagonal.
    Green dash-dotted: $\RDV > \RDT + 20\mm$.
    All accepted signal lies above this floor and to the right of the DT
    threshold, in the region uniquely covered by SR-DT4DV.
    Red dotted: blind-zone edges; faint grey dotted: CMS pixel-layer radii.}
  \label{fig:blindspot}
\end{figure}

Throughout, ``kinship'' denotes the matching of a reconstructed DT--DV
pair to a common generated $\Ntwo$ in signal Monte Carlo. It is used
\emph{solely} as a signal-MC diagnostic---to identify the correct decay
chain when validating the back-pointing resolution and when producing
the signal distributions of Fig.~\ref{fig:blindspot}---and is \emph{not}
an analysis requirement. The signal-region selection is defined entirely
by the reconstruction-level trigger, disappearing-track, displaced-vertex,
4D-timing, and geometric back-pointing cuts above; the physical DT--DV
correlation is enforced at detector level by the pointing cut (step~5),
not by truth information.

The interval $[50,700]\mm$ is the \emph{generic} displaced-vertex
reconstruction fiducial (inner to outer tracker); within SR-DT4DV the
nested requirement $\RDV>\RDT+20\mm$ with $\RDT\ge160\mm$ forces
$\RDV\ge180\mm$, so the effective DV range is $[180,700]\mm$ and the
$[50,180]\mm$ strip is never populated. This changes no yield: the
$\RDV>\RDT+20\mm$ cut is applied explicitly in Table~\ref{tab:cutflow}.

Table~\ref{tab:cutflow} presents the cutflow for the benchmark
scenario at $(\ctauH, \ctauN) = (300\mm, 189\mm)$,
obtained from $5\times10^4$ fully simulated events.
The DT step reflects the conservative requirement
$\RDT \geq 160\mm$: for $\ctauH = 300\mm$ the mean lab-frame
decay radius $\langle\RDT\rangle \approx 210\mm$ lies inside the
window, so the $[160,300]\mm$ requirement retains a large fraction
of $\etapm$ decays, yielding
$\varepsilon_{\rm DT} = 21.4\%$.
The 4D timing cut~(step~4) strongly suppresses the time-prompt
material-conversion background (additionally removed by a material-map
veto; see below); the pointing condition~(step~5) retains ${\sim}100\%$
of signal, as the reconstructed DT-to-DV vector aligns with the
disappearing-track direction to ${\approx}2.5\,\text{mrad}$, far inside
the $0.1\,\text{rad}$ window.
The $\mu^+\mu^-$ excess over $e^+e^-$ ($278$ vs $76$)
reflects the Path~B displaced-muon trigger bias, as expected.
The SM background quoted in the table is the
conservative upper bound derived in the Backgrounds paragraph
below, dominated by the pileup combinatoric sub-case~B;
it is not a raw \textsc{Delphes} MC event count.

\begin{table}[ht]
\caption{Cutflow for the benchmark point
  $\mNtwo = 150\GeV$, $\mNone = 100\GeV$,
  $\mHpm = 150.2\GeV$,
  $({\ctauH}, {\ctauN}) = (300\mm, 189\mm)$
  ($50\,000$ generated events).
  All efficiencies are cumulative.
  The final yield is broken down by displaced-vertex lepton
  content ($e^+e^-$, $\mu^+\mu^-$, or a single reconstructed
  lepton; these sum to the total).
  The SM background row is the analytic upper bound derived in
  Sect.~\ref{sec:analysis}, not a Monte-Carlo event count.}
\label{tab:cutflow}
{\small
\begin{tabular}{lrr}
\hline\noalign{\smallskip}
Selection & $N$ & $\varepsilon$ [\%] \\
\noalign{\smallskip}\hline\noalign{\smallskip}
Generated                                      & 50\,000 & 100.00 \\
${\geq}1$ DT; $\RDT\!\in\![160,300]\mm$       & 10\,681 &  21.36 \\
Trigger (A $\cup$ B)                           &  5\,629 &  11.26 \\
$\ell^\pm$ DV (disp.\ $+$ lepton $+$ pointing) &  1\,412 &   2.82 \\
\quad $+\,\RDV\!\le\!700\mm$,
  $m_{\ell\ell}\!<\!48\GeV$                    &  1\,278 &   2.56 \\
$\RDV > \RDT + 20\mm$ [nested]                 &    893 &   1.79 \\
$\dt > 200\ps$ (displaced vertex)             &    699 &   1.40 \\
$\MET > 50\GeV$ [SR-DT4DV]                     &    590 &   1.18 \\
\noalign{\smallskip}\hline\noalign{\smallskip}
\textbf{Final yield}            & \textbf{590} & \textbf{1.18} \\
\quad $e^+e^-$ DV               & 76             & --- \\
\quad $\mu^+\mu^-$ DV            & 278            & --- \\
\quad single-lepton DV          & 236            & --- \\
\noalign{\smallskip}\hline\noalign{\smallskip}
SM background                   & $\lesssim 10^{-3}$ & --- \\
\noalign{\smallskip}\hline
\end{tabular}
}
\end{table}

Figure~\ref{fig:blindspot} shows the signal distribution in the
$(\RDT, \RDV)$ plane. (In signal Monte Carlo the generated $\Ntwo$
chain is used only to select the correct DT--DV pair for this
diagnostic plot, as noted above; it plays no role in the selection.)
The DV-only blind zone (blue shading, $\RDT < 160\mm$) corresponds
to events where $\etapm$ decays before crossing all four barrel
pixel layers, leaving an unreconstructable DT stub.
Although a displaced vertex can in principle be reconstructed down to
$\RDV\simeq50\mm$ (the generic tracker fiducial), the nested
requirement $\RDV>\RDT+20\mm$ with $\RDT\ge160\mm$ sets an effective
lower edge $\RDV\ge180\mm$ inside SR-DT4DV; the $\RDV<180\mm$ strip is
therefore never populated.
Accepted signal events populate the unshaded gap
($\RDT \in [160,300]\mm$, $180\mm\le\RDV<700\mm$,
$\RDV > \RDT + 20\mm$), uniquely covered by SR-DT4DV.
(The scatter shows all truth decay radii, so points with $\RDV>700\mm$
correspond to $\Ntwo$ decays rejected by the upper $\RDV$ requirement.)

The double-delay observable $\langle\dt\rangle$ accumulates
contributions from both the $\etapm$ flight (PV$\to$DT) and the
subsequent $\Ntwo$ flight (DT$\to$DV), as given by
Eq.~(\ref{eq:deltat}).
The resulting signal arrival-time residual lies well above the
$\dt \approx 0$ expectation for prompt SM processes, so that the
$\dt > 200\ps$ requirement retains the bulk of the signal while
strongly suppressing the prompt-timed background; the per-step
yields are listed in Table~\ref{tab:cutflow}.


At the benchmark
($\mNtwo\!=\!150\GeV$, $\mHpm\!=\!150.2\GeV$,
$\ctauH\!=\!300\mm$, $\ctauN\!=\!189\mm$),
the SR-DT4DV efficiency with $\RDT \geq 160\mm$ is
$\varepsilon = 1.18\%$.
Using the fiducial NLO cross section
$\sigma_{\rm fid}^{\rm NLO} = 34.24\fb$
($= K \sigma_{\rm fid}^{\rm LO}$,
$K=1.25$~\cite{Fuks:2012qx,Fuks:2013lya},
PDF4LHC21\_40~\cite{PDF4LHC21:2022mnw}),
the expected HL-LHC signal yield is
$N_s = \varepsilon\,\sigma_{\rm fid}^{\rm NLO}\,\mathcal{L}
= 0.0118 \times 34.24\fb \times 3000\ifb \approx 1212\,^{+14\%}_{-11\%}$ events (scale$\,\oplus\,$PDF),
with an expected SM background $B\sim10^{-5}$--$10^{-3}$ events
(see background discussion below). The reach is insensitive to this
value: a $5\sigma$ discovery requires only $N_s\simeq2,4,8,20$ for
$B=10^{-3},0.1,1,10$ events respectively, so the benchmark $N_s=1212$
yields a $\gtrsim96\sigma$ median significance for any $B\lesssim10$.
The $\pm14\%$ normalisation uncertainty on $N_s$ from scale and PDF
variations is negligible against this four-order-of-magnitude margin.

Figure~\ref{fig:2d_contour} shows the 95\%\,CL exclusion
and $5\sigma$ discovery reach in the $(\ctauH, \ctauN)$ plane
at $3000\ifb$, from the full 56-point efficiency scan.
The colour map encodes $N_s$ on a logarithmic scale.
The 95\%\,(68\%)\,CL exclusion boundaries ($N_s = 2.996$ and $1.14$,
dark blue solid/dashed) follow from the CLs
method~\cite{Read:2002hq} in the zero-observed-events limit,
$\mathrm{CL}_s = e^{-N_s} < 0.05$, and are \emph{independent} of $B$.
The $5\sigma$ discovery contour does depend on $B$; the dark-red band
spans $B = 10^{-3}$--$10$ events, with the conservative $N_s = 6$
threshold (valid for $B\lesssim0.2$) shown as the reference line.
The 68\%\,CL boundary ($N_s = 1.14$, dark blue dashed)
is also shown.
Of the 56 scanned points, the 40 with non-zero DT acceptance
($\ctauH\gtrsim15\mm$) all exceed both the $5\sigma$ and 95\%\,CL
thresholds; the remaining 16 ($\ctauH=1$ and $5\mm$) fall in the
DV-only blind region, where the $\etapm$ decays before traversing the
$[160,300]\mm$ DT window. The effective displaced-vertex floor
$\RDV\ge180\mm$ discussed above does not alter this $40/16$ split, nor
any tabulated yield: all accepted events already satisfy
$\RDV>\RDT+20\mm\ge180\mm$, and the $16$ blind points are excluded
solely by the DT window. Because the signal yield rises steeply across
the boundary, the $5\sigma$ band and the exclusion line nearly coincide
and shift only mildly as $B$ varies over the range considered.
Peak sensitivity is reached at
$(\ctauH, \ctauN) = (150\mm, 60\mm)$ with $N_s \approx 1635$ events,
while the physical benchmark $(300\mm, 189\mm)$ yields $N_s \approx 1212$.

The nested DT$+$DV strategy covers the central region
$15\mm \lesssim \ctauH \lesssim 300\mm$,
$10\mm \lesssim \ctauN \lesssim 1000\mm$;
the $\ctauN$ span probes neutrino-sector Yukawa couplings
$6\times10^{-4} \lesssim y_1 \lesssim 4\times10^{-3}$
of the scale that generates sub-eV neutrino masses.
No existing search exploits the correlated DT$+$DV kinematic
chain simultaneously; individual DT and DV searches may retain
partial, unoptimised acceptance for subsets of this topology,
but none targets the nested cascade studied here.
Standalone DT searches~\cite{ATLAS:2022rme,CMS:2020atg,ATLAS:2026hnb,CMS:2023mny}
require the downstream $\Ntwo$ to be effectively stable or
invisible, a condition not guaranteed across our scan;
searches for nearly-degenerate higgsinos via low-$p_T$ displaced
tracks~\cite{ATLAS:2024umc} target a different kinematic regime
($\Delta M \sim$~few~GeV) with different trigger requirements.
Standalone DV searches for displaced leptons and
MET~\cite{ATLAS:2023oti,CMS:2021kdm,ATLAS:2026riv,ATLAS:2026gcd,%
CMS:2024trg,ATLAS:2024vnc,CMS:2025qkk}
would have partial acceptance for the $\Ntwo$ DV in isolation,
but none imposes the upstream DT requirement that defines SR-DT4DV
and provides its strong, correlated background rejection.
The present analysis is specifically designed for the
compressed-spectrum regime where both $\etapm$ and $\Ntwo$
are simultaneously long-lived.

\begin{figure}[ht]
  \centering
  \includegraphics[width=0.9\columnwidth]{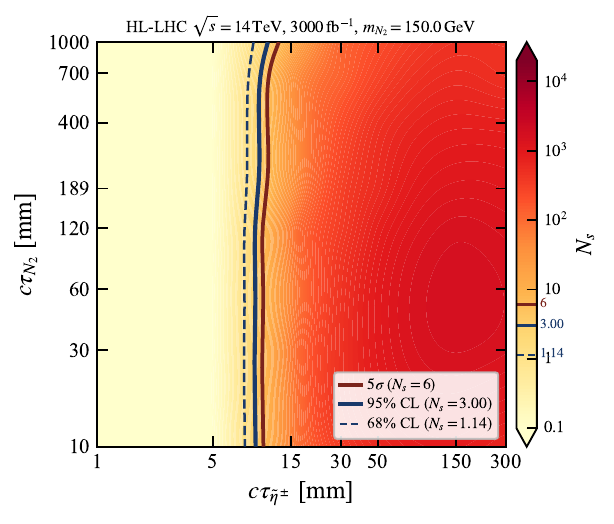}
  \caption{Sensitivity in the $({\ctauH}, {\ctauN})$ plane at
    $\sqrt{s}=14\,\text{TeV}$, $3000\ifb$, $\mNtwo=150\GeV$.
    Mass assumptions: $\mHpm = 150.2\GeV$, $\mNone = 100\GeV$,
    $\DMcomp = 200\MeV$, $\DMopen = 50\GeV$.
    Colour map: $N_s$ (logarithmic scale).
    Dark red band: $5\sigma$ discovery contour for $B = 10^{-3}$--$10$
    events (reference line $N_s = 6$, valid for $B\lesssim0.2$).
    Dark blue solid/dashed: 95\%\,(68\%)\,CL exclusion
    ($N_s = 2.996$ and $1.14$, CLs zero-observed-events limit,
    independent of $B$).
    The pale region at $\ctauH \lesssim 10\mm$ reflects
    the DV-only insensitive zone where $\etapm$ decays
    before crossing all four barrel pixel layers.}
  \label{fig:2d_contour}
\end{figure}

\textbf{Backgrounds and discovery reach.}
We first establish that the discovery reach is \emph{robust against the
background level}, then estimate where the background is expected to
sit. Treating the expected SM background $B$ as a parameter, the median
signal yield required for a $5\sigma$ discovery rises only slowly with
$B$: from $N_s\simeq2$ at $B=10^{-3}$ to $N_s\simeq4$ at $B=0.1$,
$N_s\simeq8$ at $B=1$, and $N_s\simeq20$ at $B=10$ events
(Asimov median significance~\cite{Cowan:2010js}).  Equivalently, in
frequentist terms the observed-count threshold for $5\sigma$ is
$n\geq3,\,5,\,10,\,30$ for $B=10^{-3},\,0.1,\,1,\,10$.  At the expected
background ($B\lesssim10^{-3}$; see the cross-check below) a $5\sigma$
discovery requires only $N_s\lesssim2$, so all 40 scanned points with
non-zero DT acceptance are discovered, the weakest being
$N_s\approx12$ at $(\ctauH,\ctauN)=(15,1000)\mm$.  Even allowing the
background to be as large as $B=10$ events---four orders of magnitude
above the expected level---$39$ of these $40$ points retain a $5\sigma$
discovery ($N_s\gtrsim20$); only the longest-lifetime corner drops below,
and even there a $5\sigma$ discovery survives for any $B\lesssim1$.  At
the benchmark ($N_s\approx1212$) the median significance stays
$Z\gtrsim30\sigma$ even in the extreme case $B=N_s$ (signal-to-background
unity).  The conclusion of this study is therefore insensitive to the
precise value of $B$ over more than four orders of magnitude; the role of
the background estimate below is only to locate $B$ within this robust
range, not to underpin the discovery.

\emph{Monte-Carlo cross-check.}
A dedicated background simulation of six SM processes
($b\bar b$, $c\bar c$, $t\bar t$, $Z\to\tau\tau$, diboson, Drell--Yan$+$jet),
${\sim}2\times10^{5}$ events each, passed through the identical
reconstruction-level selection (\texttt{--no-fiducial}), yields
\emph{zero} surviving events: prompt SM backgrounds are removed at the
trigger and displaced-vertex stages before the 4D-timing requirement is
even reached.  With these statistics the corresponding $95\%$\,CL
\emph{statistical} upper limits are large
($10^{2}$--$10^{8}$ events, set by the available MC sample size, not by
the physical background); reducing them to the $\lesssim1$-event level by
brute force would require ${\sim}10^{13}$ simulated events.  This
quantitatively confirms that a Monte-Carlo background estimate is not
feasible here and that the residual background must be assessed
analytically and, ultimately, with data-driven techniques --- exactly as
in all existing ATLAS and CMS DT/DV
searches~\cite{ATLAS:2022rme,CMS:2020atg,ATLAS:2026hnb,CMS:2023mny,ATLAS:2026riv}.
We stress that these limits are \emph{not} a background prediction.

\emph{Analytic estimate of the residual background.}
The dominant arrival-time discriminant is the $\Delta t>200\,\ps$ cut.
For any $\beta\approx1$ process the time residual is set by the MTD
resolution, $\Delta t\sim\mathcal{N}(0,\sigma_t)$ with $\sigma_t=30\,\ps$,
so that
\begin{equation}
  P(\Delta t > 200\,\ps\,|\,\beta=1)
  = \tfrac12\,\mathrm{erfc}\!\left(\frac{200\,\ps}{30\,\ps\sqrt2}\right)
  \approx 1.3\times10^{-11}
\label{eq:timing_tail}
\end{equation}
within the Gaussian core; requiring both displaced leptons gives
$P_{\rm both}\approx1.7\times10^{-22}$.  We emphasise that
Eq.~(\ref{eq:timing_tail}) describes the \emph{Gaussian} response only:
the realistic timing floor is set by \emph{non-Gaussian} tails and
wrong-primary-vertex (wrong-PV) association, which are not captured by a
$6.7\sigma$ extrapolation and cannot be modelled without full
pileup-overlay simulation.  We therefore do not rely on the resulting
sub-$10^{-19}$ numbers as literal bounds; the prompt sources below are
simply \emph{negligible} relative to the dominant pileup combinatoric
term, which is itself controlled by this same non-Gaussian floor.

\emph{(i) Material conversions ($\gamma\to e^+e^-$) and
(ii) heavy-flavour / fake DVs.}
Photon conversions are \emph{time}-prompt ($\beta\approx1$): the
converting photon travels at $c$, so the $e^+e^-$ pair carries
essentially zero arrival-time delay and is removed by the
$\dt>200\,\ps$ requirement.  Spatially, however, conversions are
\emph{not} prompt---they occur on the active and passive detector
material and appear as genuinely displaced vertices localised on the
material layers.  Displaced-vertex analyses suppress this component
with a material-map veto that removes vertices coincident with known
material~\cite{ATLAS:2023oti}, which we assume is applied here; the
converted pair is further rejected by the signal selection because it
has $m_{ee}\approx0$ (well below the $\Ntwo\to\None\ell^+\ell^-$
dilepton spectrum) and does not back-point to a disappearing-track
endpoint.  Heavy-flavour leptons are likewise time-prompt, and
$b$-hadron decays give $R_{\rm DV}\lesssim5\,\mm$, far below the
$50\,\mm$ requirement.  Within the Gaussian timing model both are
suppressed by $P_{\rm both}\approx10^{-22}$; even allowing a generous
non-Gaussian tail at the $10^{-3}$--$10^{-4}$ level per object, and
combined with the material veto and the mass/pointing requirements,
these sources remain $B\ll10^{-3}$ events.

\emph{(iii) Pileup combinatorics (dominant).}
This is the only background that approaches the relevant scale and the
one \textsc{Delphes~3} cannot estimate, as it requires pileup track-stub
fake rates and MTD 4D vertex-time assignment at
$\langle\mu\rangle=200$.  We anchor to the ATLAS Run~2
disappearing-track search~\cite{ATLAS:2026hnb}, whose data-driven
fake-tracklet estimate is $0.33\pm0.14$ events in $137\,\ifb$ at
$\langle\mu\rangle=34$ in a 4-layer,
$p_T^{\rm tracklet}>60\,\GeV$, $E_T^{\rm miss}>300\,\GeV$ selection.
SR-DT4DV adds three requirements with no Run~2 analogue --- a correlated
displaced dilepton DV ($P_{\rm DV}\approx1.6\times10^{-2}$), spatial
back-pointing $\Delta\phi,\Delta\theta<0.1\,\text{rad}$
($P_{\rm point}\approx(0.1)^2/\pi\approx3\times10^{-3}$), and the
$\Delta t>200\,\ps$ timing cut.  For correctly vertex-assigned pileup the
timing factor is $P_{\rm both}\approx10^{-22}$; the residual is set
entirely by the wrong-PV population, whose rate at
$\langle\mu\rangle=200$ for this displaced, same-vertex topology is an
\emph{extrapolation} of the ${\lesssim}1\%$ figure measured in less
demanding conditions~\cite{CMSCollaboration:2667167}.  Scaling the ATLAS
baseline to HL-LHC ($\,{\sim}43$ events, conservatively assuming linear
pileup growth) and folding in these factors with a wrong-PV timing pass
probability $P_{\rm timing}\lesssim10^{-2}$ gives a central estimate
\begin{equation}
  B_{\rm PU}\ \sim\ 43 \times 1.6\times10^{-2}\times3\times10^{-3}
            \times10^{-2}\ \approx\ 2\times10^{-5}\ \text{events}.
\label{eq:bkg_pileup}
\end{equation}
The wrong-PV rate and the pileup-scaling law are the leading
uncertainties; even allowing them to be underestimated by two orders of
magnitude raises this to $B_{\rm PU}\sim\text{few}\times10^{-3}$
events --- still deep inside the robust range established above, where
the discovery is unaffected.  A definitive number requires a data-driven
control region (an anti-pointing sideband
$0.1<\Delta\phi_{\rm DT\to DV}<0.5\,\text{rad}$) or full pileup-overlay
simulation with MTD 4D tracking, both beyond the scope of this
phenomenological study.

\emph{(iv) Cosmic rays and beam-induced background.}
Cosmic muons are rejected by the ISR-jet--from--PV requirement and the
collision-timing window ($|t_{\rm arr}|<5\,\text{ns}$); beam halo
($|\eta|\gg2.1$) is removed by collision timing and ML-based vetoes
developed for ATLAS DT/DV
searches~\cite{ATLAS:BIB2023,ATLAS:BIB2024}.  Both are $B\ll10^{-2}$
events.

\textbf{Summary.}
The residual SM background is expected at the
$B\sim10^{-5}$--$10^{-3}$ level, dominated by pileup combinatorics and
ultimately requiring HL-LHC data-driven validation.  Crucially, the
discovery reach reported below does not depend on this expectation:
it holds for any $B\lesssim10$ events, a margin of four orders of
magnitude (Fig.~\ref{fig:2d_contour} and Table~\ref{tab:backgrounds}).

\begin{table*}[ht]
\caption{SM background sources in SR-DT4DV at $3000\ifb$
  ($\sqrt{s}=14\,\text{TeV}$, $\langle\mu\rangle=200$), and the robustness of
  the reach. Prompt sources (i)--(ii) are removed by the displaced and
  timing requirements; their residual is bounded by the non-Gaussian MTD
  timing tail rather than the Gaussian extrapolation of
  Eq.~(\ref{eq:timing_tail}). The dominant source is pileup combinatorics
  (iii), estimated from an ATLAS Run~2 data-driven anchor. The final row
  states the key result: the $5\sigma$ discovery is retained for any
  background up to $B\sim10$ events.}
\label{tab:backgrounds}
{\small
\begin{tabular}{p{3.0cm}p{5.4cm}p{2.8cm}p{4.4cm}}
\hline\noalign{\smallskip}
Source & Dominant suppression & Estimate & Method / limitations \\
\noalign{\smallskip}\hline\noalign{\smallskip}
(i)~$\gamma\!\to\!e^+e^-$, (ii)~HF / fake DV
  & \emph{Time}-prompt ($\beta\!\approx\!1$): $\Delta t>200\ps$;
    conversions are spatially displaced on material, removed by a
    material-map veto; $m_{ee}\!\approx\!0$ and back-pointing also reject
  & $\ll10^{-3}$ events
  & Analytic; MC infeasible; bounded by non-Gaussian MTD tail
    (not the $6.7\sigma$ Gaussian value) \\[4pt]
(iii)~Pileup stub $+$ pileup DV \emph{(dominant)}
  & Correlated DV $\times$ pointing $\times$ wrong-PV timing
    ($\approx5\times10^{-7}$ beyond the Run~2 anchor)
  & ${\sim}2\times10^{-5}$, conservatively
    $\lesssim\text{few}\times10^{-3}$
  & ATLAS Run~2 data-driven anchor~\cite{ATLAS:2026hnb} scaled to HL-LHC;
    \textsc{Delphes} cannot model pileup overlay or MTD 4D tracking;
    \emph{requires data-driven control region} \\[4pt]
(iv)~Cosmic / beam halo
  & ISR-jet $+$ timing window $+$ ML veto
  & $\ll10^{-2}$ events
  & Qualitative, following ATLAS practice~\cite{ATLAS:BIB2023,ATLAS:BIB2024} \\
\noalign{\smallskip}\hline\noalign{\smallskip}
\textbf{MC cross-check}
  & Six SM processes, ${\sim}2\times10^5$ events each, full reco selection
  & \textbf{0 survivors}
  & Confirms prompt backgrounds removed; $95\%$\,CL \emph{statistical}
    limits $10^{2}$--$10^{8}$ are MC-stat-limited, \emph{not} a
    background prediction \\[4pt]
\textbf{Robustness}
  & Median $5\sigma$ needs $N_s\!\simeq\!2/4/8/20$ at
    $B\!=\!10^{-3}/0.1/1/10$
  & \boldmath$5\sigma$ \textbf{at expected} $B$; \textbf{robust to}
    $B\!\lesssim\!10$
  & $40/56$ points discovered at expected $B$; weakest
    $N_s\!\approx\!12$ stays $5\sigma$ for $B\!\lesssim\!1$;
    benchmark $N_s\!=\!1212\!\Rightarrow\!Z\!\gtrsim\!30\sigma$ at $B\!=\!N_s$ \\
\noalign{\smallskip}\hline
\end{tabular}
}
\end{table*}

\section{Conclusion}
\label{sec:conclusion}

We have proposed a search strategy for the compressed-spectrum
2IDM3N scotogenic model at the HL-LHC exploiting a
\emph{nested disappearing-track plus displaced-vertex topology},
in which both $\etapm$ and $\Ntwo$ are simultaneously long-lived.
The signal region SR-DT4DV requires a disappearing $\etapm$ track
at $\RDT\in[160,300]\mm$ (four barrel pixel layers), a downstream
$\Ntwo\to\None\,\ell^+\ell^-$ displaced vertex at
$\RDV>\RDT+20\mm$, and a 4D timing cut $\dt>200\ps$ that, together
with a material-map veto, strongly suppresses material-conversion
backgrounds~\cite{Liu:2018wte,ATLAS:2023oti};
a spatial pointing condition ($\delta\theta\lesssim3\,\text{mrad}$)
provides additional suppression with no SM analogue.
The total background is $B\lesssim10^{-3}$ events, dominated
by pileup combinatorics and requiring HL-LHC data-driven
validation (Sect.~\ref{sec:analysis}).

Full simulation at the benchmark ($\mNtwo=150\GeV$,
$\ctauH=300\mm$, $\ctauN=189\mm$) yields $\varepsilon=1.18\%$
and $N_s\approx1212$ events at $3000\ifb$.
Across the 56-point lifetime scan, the 40 points with non-zero DT
acceptance ($\ctauH\gtrsim15\mm$) all exceed both the $5\sigma$
discovery reach and the 95\%\,CL exclusion limit; the 16 points at
$\ctauH=1$ and $5\mm$ lie in the DV-only blind region.  Because the
discovery is robust against the background level (retained for any
$B\lesssim10$ events, Sect.~\ref{sec:analysis}) and the signal yield
rises steeply across the boundary, the discovery and exclusion contours
nearly coincide. Peak sensitivity is reached at
$(\ctauH,\ctauN)=(150\mm,60\mm)$ with $N_s\approx1635$ events;
the physical benchmark $(300\mm,189\mm)$ yields $N_s\approx1212$.
The covered region
$15\mm\lesssim\ctauH\lesssim300\mm$,
$10\mm\lesssim\ctauN\lesssim1000\mm$
probes neutrino-sector couplings
$6\times10^{-4}\lesssim y_1\lesssim4\times10^{-3}$
through the $\Ntwo$ decay length, the Yukawa range associated
with sub-eV neutrino masses,
and is not simultaneously targeted by any existing ATLAS or
CMS search.
This work provides a concrete roadmap for including the
scotogenic nested cascade in the HL-LHC Run~4 search programme.

\medskip
\noindent\textbf{Acknowledgement.}\quad
The author thanks colleagues at the Institute of High Energy
Physics for stimulating discussions.
This work is supported by the Internal Research Fund of the
Institute of High Energy Physics, Chinese Academy of Sciences.

\medskip
\noindent\textbf{Data availability statement.}\quad
This manuscript has no associated data.

\medskip
\noindent\textbf{Code availability statement.}\quad
This manuscript has no associated code/software.

\bibliographystyle{elsarticle-num}
\bibliography{references}

\end{document}